\documentstyle[aas2pp4,psfig]{article}

\slugcomment{Revised version submitted: July 8, 1999}
\lefthead{Smale}
\righthead{No Dips in X1254$-$690}

\def\sourcea{X1254$-$690}

\begin{document}

\title{A Cessation of X-ray Dipping Activity in X1254$-$690}

\author{Alan P. Smale\altaffilmark{1,2}}
\affil{Laboratory for High Energy Astrophysics,
Code 662, NASA/Goddard Space Flight Center, Greenbelt, MD 20771}

\author{Stefanie Wachter}
\affil{Cerro Tololo Interamerican Observatory, Casilla 603, La Serena,
Chile}

\altaffiltext{1}{Also Universities Space Research Association.} 

\altaffiltext{2}{ Visiting Astronomer, Cerro Tololo Interamerican
Observatory, National Optical Astronomy Observatories, operated by
AURA, Inc. under cooperative agreement with the NSF.}

\begin{abstract}

We present results from a campaign of simultaneous X-ray and optical
observations of \sourcea\ conducted using the {\sl Rossi} X-ray Timing
Explorer and the CTIO 1.5m telescope. We find that the
usually-observed deep X-ray dipping is not seen during the times of our
observations, with an upper limit of $\sim$2\% on any X-ray orbital
variation, and that the mean optical variability has declined in amplitude
from $\Delta V$=0.40$\pm$0.02 mag to 0.28$\pm$0.01 mag.  These
findings indicate that the vertical structure on the disk edge
associated with the impact point of the accretion stream has decreased
in angular size from 17--25$^o$ to $<$10$^o$, and support the
suggestions of previous modeling work that the bulge provides 35--40\%
of the contribution to the overall optical modulation. The average
optical and X-ray brightnesses are comparable to their values during
dipping episodes, indicating that the mean $\dot M$ and disk radius
remain unchanged.

\end{abstract}

\keywords{accretion, accretion disks --- stars: individual
(X1254$-$690) --- stars: neutron --- stars: binaries: close ---
X-rays: stars}

\section{Introduction}

In X-ray dipping sources, accretion disk structure extends
vertically above the plane of the binary and periodically blocks the
line of sight to the central compact object. This structure is
believed to be a bulge in the disk edge, associated with the impact
point of the accretion stream from the secondary star. Because of this
azimuthal material, the X-ray light curves of these low-mass binaries
contain deep, irregularly-shaped dips associated with large increases
in absorption, that recur on the orbital period of the
system. Currently we know of $\sim$10 X-ray dippers.

\sourcea\ has an orbital period of 3.88 hrs, and generally
displays dips of up to 95\% of the 1--10 keV flux lasting $\sim$0.8 hr
per cycle (Courvoisier et al.\ 1986). Its counterpart, GR~Mus, has a
mean optical brightness of $V$=19.0 mag, and shows a broad optical
modulation probably due to the changing visibility of the heated face
of the secondary star in the system, with an additional contribution
from the X-ray-heated bulge (Motch et al.\ 1987, hereafter MEA). A
short (5-hr) section of simultaneous X-ray and optical coverage showed
that the optical minimum occurs 0.15 in phase after the center of the
X-ray dips (MEA), supporting the system geometry described above.

Modeling of the observed variability leads to constraints on the
source inclination of 65$^\circ$--73$^\circ$, and on the distance of
8--15 kpc (MEA), and makes the prediction that the shape of the
optical light curve should be correlated with the length of the X-ray
dips, since both depend upon the angular extent of the accretion disk
structure.  In an attempt to confirm the linkage between the observed
optical and X-ray characteristics in LMXBs, we undertook a program of
simultaneous observations of \sourcea\ using the {\sl Rossi} X-ray
Timing Explorer (RXTE) and the facilities at CTIO. 

\section{Observations}

Observations of the optical counterpart of \sourcea\ were performed at
the CTIO 1.5m telescope on 1997 April 28 -- May 1, using the Tek2k CCD
with a pixel scale of 0.24'' pix$^{-1}$. We obtained a total of 28
hours of CCD photometry, with typical exposure times of 600 sec in $V$
and 900 sec in $B$.  Overscan and bias corrections were made for each
CCD image with the {\it quadproc} task at CTIO to deal with the
4-amplifier readout. The data were flat-fielded in the standard manner
using IRAF, and photometry was performed by point spread function
fitting using DAOPHOT II (Stetson 1993). The instrumental magnitudes
were transformed to the standard system through observations of
several Landolt standard star fields (Landolt 1992). The systematic
error (from the transformation to the standard system) in these
optical magnitudes is $\pm$0.10 mag, although the internal intrinsic
1$\sigma$ error in the relative photometry is estimated to be
$\pm$0.02 mag, based on the rms scatter in the light curves of
comparison stars of similar brightness.

Simultaneous X-ray data from \sourcea\ were obtained using the RXTE
satellite. The data presented here were collected by the PCA
instrument using the Standard 2 and Good Xenon configurations, with
time resolutions of 16 sec and $<1\mu$sec respectively, and full
spectral resolution.  The scheduling flexibility of RXTE enabled us to
achieve an impressive degree of simultaneity with our ground-based
observing program, with 70 ksec of total coverage during our optical
run. An additional section of X-ray data lasting 10 ksec was obtained
on 1997 May 5. Background subtraction was performed using
modeling based on the instantaneous high-energy particle flux as
measured by the six-fold anticoincidence counters in the PCA, along
with minor contributions from activation during SAA passages and the
cosmic X-ray background.

\begin{figure}[htb]
\figurenum{1a}
\begin{center}
\begin{tabular}{c}
\psfig{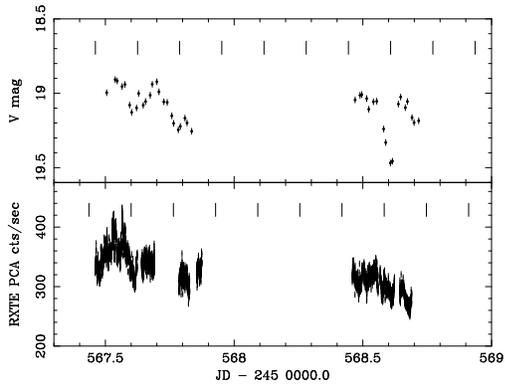}
\end{tabular}
\caption{The optical ($V$) and X-ray (2--10 keV) light
curves of \sourcea, as observed using the RXTE PCA and the CTIO 1.5m
telescope on 1997 April 28 -- May 1. The X-ray data are plotted with a
time resolution of 16 seconds. The tick marks in the upper panel
indicate the times of optical minima; the ticks in the lower panel
were generated by subtracting 0.15$\times P_{orb}$, and represent the
expected times of X-ray dip centers, assuming the standard system
geometry.
\label{x1254fig1a}}
\end{center}
\end{figure}

\begin{figure}[htb]
\figurenum{1b}
\begin{center}
\begin{tabular}{c}
\psfig{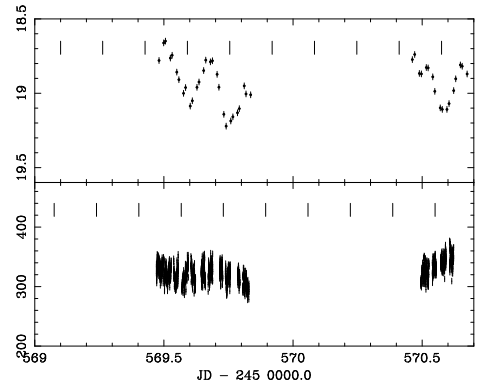}
\end{tabular}
\caption{Continuation of Figure 1.
\label{x1254fig1b}}
\end{center}
\end{figure}

The optical and X-ray light curves for these observations are shown in
Fig.\ \ref{x1254fig1a},Fig.\ \ref{x1254fig1b}. 
Tick marks are plotted in the upper panel,
indicating the expected times of optical minimum throughout the
observation, based on a folding analysis of the data and an orbital
period of $P_{orb}$=0.163890 dy.  (The ephemeris of MEA is not precise
enough to extrapolate into the epoch of the RXTE observations).  The
ticks in the lower panel were generated by subtracting 0.15$\times
P_{orb}$ from the times of optical minimum, and represent the expected
times of X-ray dip centers based on the standard LMXB system geometry.
Below, we follow the usual convention and define phase zero as the
time of optical minimum.

\section{Results}

The X-ray light curve presented in Fig.~1 bears little resemblance to
the previous behavior of the source (e.g. Courvoisier et al. 1986,
Fig.~1).  In the 2--10 keV energy range chosen for this Figure, the
observed reduction in flux during dips is generally 20--95\%, with a
dip duration of 20\% of the total cycle (Courvoisier et al.  1986); no
such dramatic activity is seen in Fig.~1, even though we have good
coverage of the expected times of all or part of five separate dips.

\begin{figure}[htb]
\figurenum{2}
\begin{center}
\begin{tabular}{c}
\psfig{figure=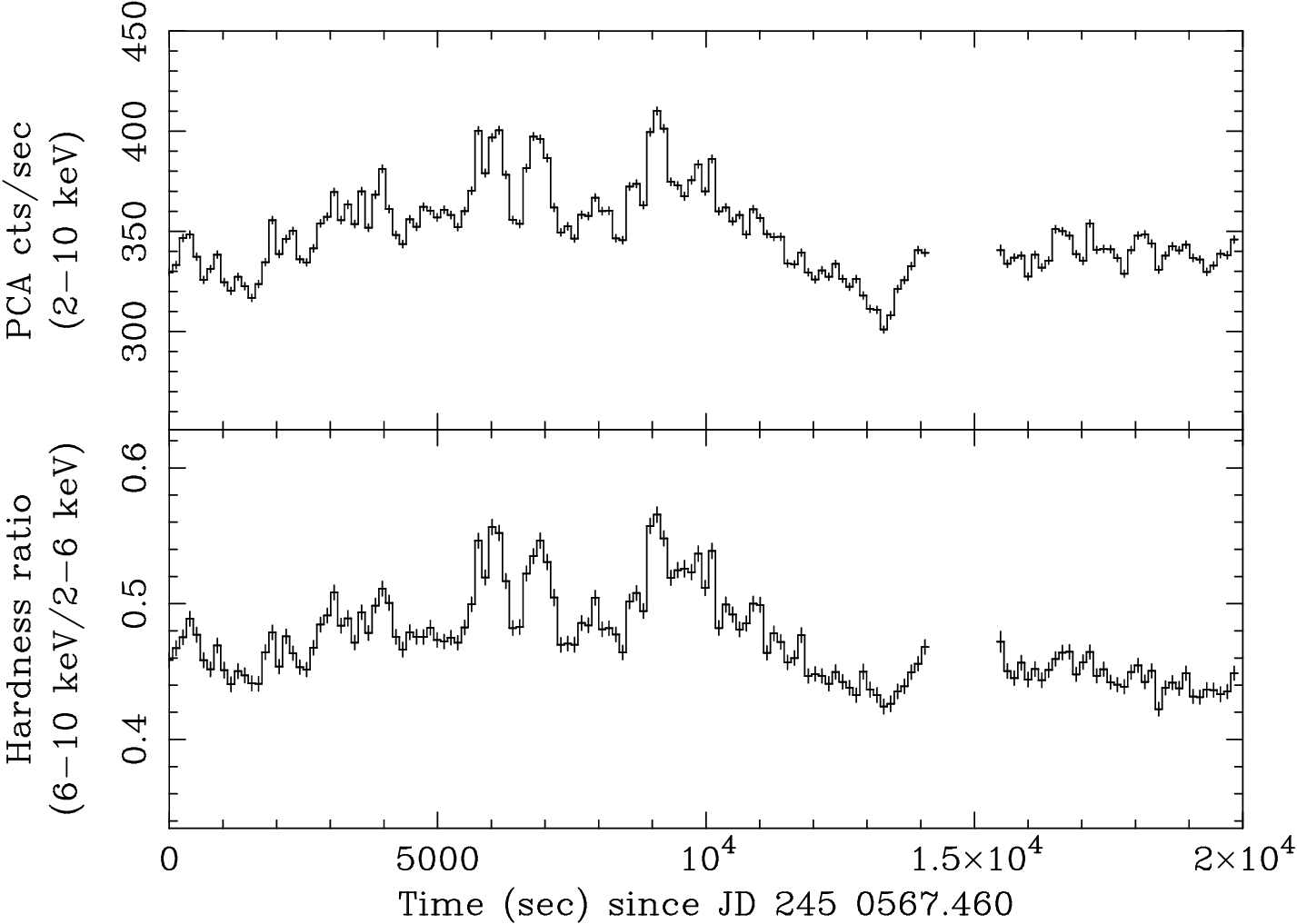,width=8cm,angle=-90}
\end{tabular}
\caption{The folded $V$ and $B$ optical light curves
for \sourcea, along with the folded X-ray light curve. The optical
curves both show quasi-sinusoidal variability on the 3.88-hr period
with full amplitudes of 0.28$\pm$0.01 and 0.30$\pm$0.01 magnitudes
respectively.  The folded X-ray light curve shows no evidence for the
usual pronounced dipping activity, and we can place an
upper limit of 2\% on the presence of a significant X-ray orbital
modulation. (During ``normal'' dipping episodes, deep $\sim$95\%
dipping is generally observed centered on phase 0.85.)  The data from each
observing night were renormalized to a common mean before performing
the folds.
\label{x1254fig2}}
\end{center}
\end{figure}

Small variations are seen in the X-ray light curve, particularly at
JD~245~0567.6. However, as we demonstrate in Fig.\ \ref{x1254fig2}, the X-ray
hardness ratio is strongly correlated with intensity during these
episodes.  This indicates that the flux reductions observed are due to
intrinsic variations in the intensity of the central source, rather
than absorption events caused by obscuration of the flux by accretion
disk material. True dipping events are associated with an increase in
low-energy absorption, and therefore an {\sl inverse} correlation
between hardness and intensity (e.g. Courvoisier et al.\ 1986; Smale
et al.\ 1988).  The 1997 May 5 data (not shown here) include an
interval spanning a sixth expected X-ray dip, but are similarly
featureless.  We conclude that there is no evidence for dipping
activity during our RXTE observations.

\begin{figure}[htb]
\figurenum{3}
\begin{center}
\begin{tabular}{c}
\psfig{figure=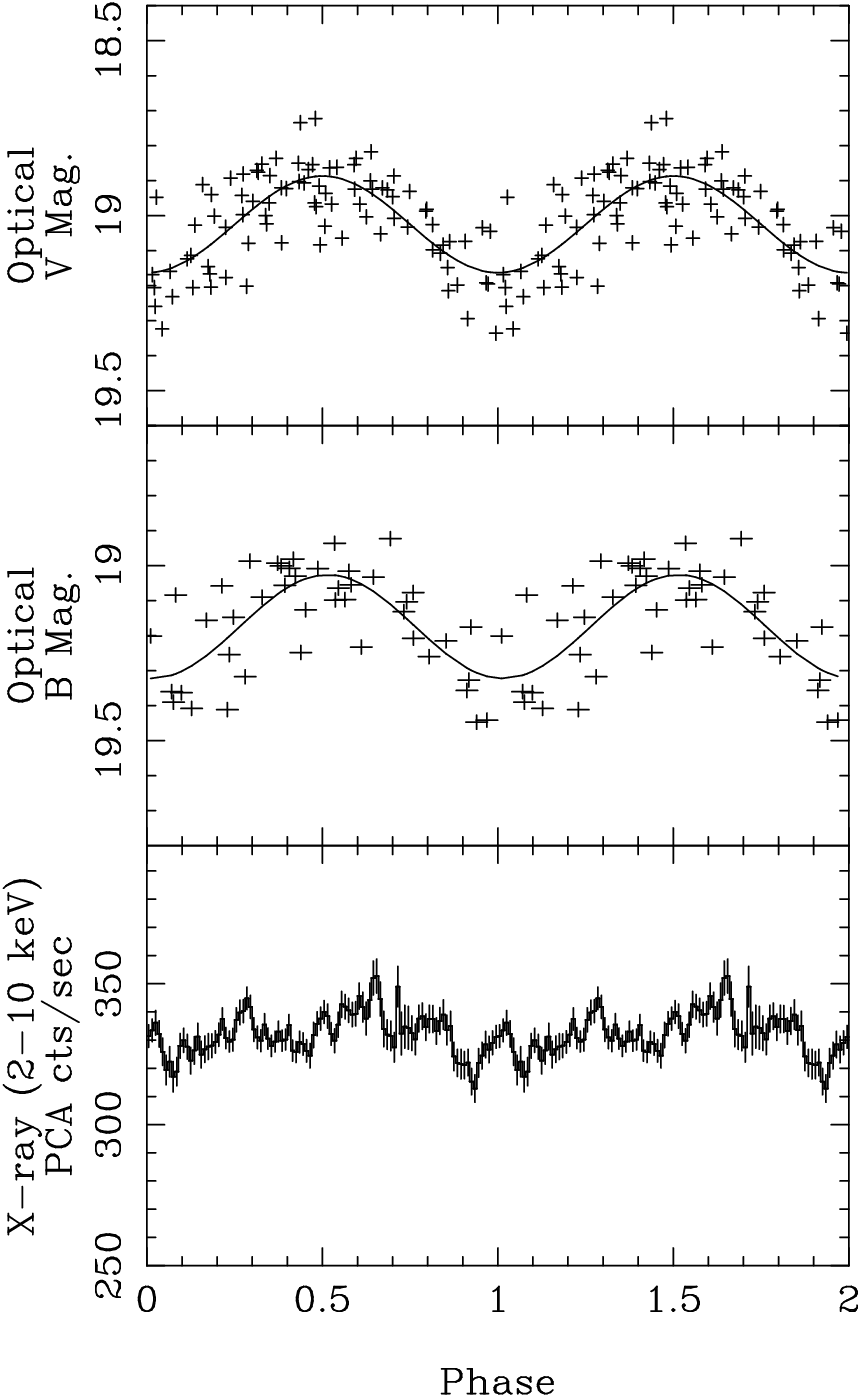,width=5.0cm}
\end{tabular}
\caption{The folded $V$ and $B$ optical light curves
for \sourcea, along with the folded X-ray light curve. The optical
curves both show quasi-sinusoidal variability on the 3.88-hr period
with full amplitudes of 0.28$\pm$0.01 and 0.30$\pm$0.01 magnitudes
respectively.  The folded X-ray light curve shows no evidence for the
usual pronounced dipping activity, and we can place an
upper limit of 2\% on the presence of a significant X-ray orbital
modulation. (During ``normal'' dipping episodes, deep $\sim$95\%
dipping is generally observed centered on phase 0.85.)  The data from each
observing night were renormalized to a common mean before performing
the folds.
\label{x1254fig3}}
\end{center}
\end{figure}

Our spectral fitting bears out this conclusion. Fitting short sections
of data accumulated before, during, and after the apparent flux
reduction episode at JD 245~0567.6 seen in the first night of RXTE
observing (Fig.~1), we find that the 2--15 keV spectra
can be well fit with a cut-off power law with photon index of
$\alpha$=0.6$\pm$0.1 and cut-off energy $E_{cut}$=4.3$\pm$0.3 keV. We
find a mean equivalent hydrogen column density of
$N_H$=4.8$\times$10$^{21}$~cm$^{-1}$, and can set an 90\% upper limit
on the increase of this $N_H$ during the flux reduction episode of
4$\times$10$^{21}$~cm$^{-1}$, two orders of magnitude less than the
$N_H \sim$5$\times$10$^{23}$~cm$^{-1}$ measured during the deep dips
of Courvoisier et al.\ (1986).

In Fig.\ \ref{x1254fig3}\ we present the folded optical and X-ray light
curves for our observations. The optical light curves both show
quasi-sinusoidal variability on the 3.88-hr period, with full
amplitudes of modulation of 0.28$\pm$0.01 mag in $V$ and 0.30$\pm$0.01
mag in $B$.  By contrast, we can place an upper limit of 2\% on the
presence of a significant X-ray orbital modulation.

We find no significant correlations between the X-ray and optical
light curves on timescales of 10$^2$--10$^5$s, using the
$z$-transformed Discrete Correlation Function of Alexander (1998). We
also detect no significant lag between the $V$ and $B$ light curves,
and no significant autocorrelations in the X-ray flux. Binning up the
folded $B$ and $V$ lightcurves in Fig.~3, we can place a
90\% upper limit of 0.03 mag on the amplitude of any orbital variation
in $B-V$. No high-frequency variations were detected in the persistent
X-ray emission, with typical upper limits of 1--3\% on the rms
amplitude of any quasi-periodic oscillations between 500--1200 Hz.

\section{Discussion}

Variations in the depth and duration of dips in X-ray binaries are
well-known. In the extreme case of the short-period (50~min) source
X1916$-$053, dips can last for 10\% to 50\% of the binary cycle and
have depths of 10\% to 100\% (Smale et al.\ 1988). These dip depths
and widths are seen to vary on a timescale much greater than the
orbital period (4--6 days, Smale et al.\ 1989, Yoshida et al.\
1995). In this example, the variations in dip properties may be due to
the flux-modulating effects of a third star in the system, a scenario
that is unlikely to be widespread among LMXBs. \sourcea\ itself shows
variability in dip depth from 95\% to 20\%, and has skipped a dip in
the past; extracting the complete light curves from the 1984 and 1985
EXOSAT observations from the HEASARC archives, we find that of the
fifteen times of expected dipping covered by the data, 11 deep dips
were observed, 3 shallow dips ($\sim$20\%), and one case where the dip
is undetectable.

However, the light curves observed in the current dataset display a
lack of dip activity unprecedented among the dip sources.  In the
EXOSAT observations of Courvoisier et al. (1986), the dip duration was
observed to be $\sim$0.8 hrs, representing 20\% of the binary
cycle. If dip activity from \sourcea\ had been nominal we would have
seen evidence of 6 complete or partial dips during our RXTE
coverage. A priori, we cannot rule out the possibility that dips
occurred only during the intervals between observations during which
RXTE was not obtaining data from \sourcea, but this seems
statistically unlikely. If a dip or a non-dip are equally likely to
occur in a given cycle, the probability of observing a dip on at least
one of our 6 opportunities is $p_{dip} >$ 98\%. In other words, even
if dips occur as infrequently as half the time, the chances are slight
that we would fail to see evidence for one during our RXTE coverage.

In reality, dips are observed from dipping sources $>$95\%
of the time. In addition, the appearance or otherwise of dips in
adjacent cycles are not statistically independent events.  It
therefore seems much more realistic to postulate that dipping has
ceased entirely during our observations. We argue below that the
optical behavior of the source provides independent corroboration of 
this assertion.

In LMXBs like \sourcea\ it is generally believed that the steady
component of the optical flux comes from the disk (e.g.\ van Paradijs
\& McClintock, 1995).  Our measured mean $V$ magnitude of 19.0 is
almost identical to the $V$ brightness observed by MEA when the source
was dipping. Also, our observed 2--10 keV X-ray flux of
7.4$\times$10$^{-10}$ erg~cm$^{-2}$~s$^{-1}$ is comparable to the
non-dip intensity observed by Courvoisier et al. (1986).  We therefore
deduce that the accretion disk radius and mean mass accretion rate are
unchanged; the disk just lacks the vertical structure usually
associated with the impact point of the accretion stream.  Detailed
modeling of the disk during active dipping cycles has shown that the
typical aperture angle of the disk in \sourcea\ is 9--13$^\circ$,
while the bulge itself extends to an azimuthal height of
17--25$^\circ$ (MEA). The inclination of the system is restricted to
65--73$^\circ$. Thus, the absence of regular dipping implies that this
bulge height has shrunk in angular size from 17--25$^o$ to $<10^o$.

We expect the varying component of the optical flux to have
contributions from both the heated face of the secondary and the
bulge. The full amplitude of the optical modulation has been
previously measured at $\Delta V$=0.40$\pm$0.02 mag (MEA), and their
subsequent modeling suggests that $\Delta V$=0.25 mag of this
originates in the heated face and $\Delta V$=0.15 mag from the
bulge. Our measured full amplitude of $\Delta V$=0.28$\pm$0.01 mag is
therefore very close to the value we would predict if the optical flux
variations were now solely caused by the variable aspect of the heated
face.

Another, more circumstantial piece of corroborating evidence for our
theory is supplied by our measured mean $B-V$ of 0.15$\pm$0.02 mag
compared to the mean $B-V$ of 0.31$\pm$0.05 mag observed when the
source was dipping (MEA). Based on evidence from similar systems such
as X1636$-$536 ($P_{orb}$=3.80 hr), we would expect the photospheric
temperature of the heated face to be $\sim$25,000--35,000~K (e.g. Smale
\& Mukai, 1988), a value which matches the estimates from
hydrodynamic modeling of irradiated companion star envelopes
(typically 20,000--40,000~K; Tavani \& London 1993). The illuminated
surfaces of X-ray-heated accretion disks should have a similar
effective temperature. However, the outer edge of the accretion disk
(including much of the outer bulge) will be screened from the direct
illumination, and should therefore be much cooler -- modeling of the
UV, optical and IR light curves of X1822$-$371 ($P_{orb}$=5.57 hr)
indicates an effective temperature of 14,500~K for these
outwards-facing regions (Mason \& C\'ordova 1982), incidentally
comparable to the 'bright spot' temperatures observed in CVs of
15,000~K (Wood et al. 1989).

So, for high inclination sources with disk edge structure such as
\sourcea\ and X1822$-$371, cooler material will generally obscure at
least part of the hotter, inner disk regions.  A disappearance or
reduction in the size of the azimuthal structure then leads to a more
direct viewing of these hot regions, and an increase in the apparent
mean emission temperature.  Although hard to model without a more
detailed knowledge of the system, the reduction we see in $B-V$ is in
the correct direction for such an increase in the mean temperature of
the overall optical emission.

Thus, the X-ray and optical datasets each {\sl independently} suggest
a shrinkage in the size of the bulge at the point where the accretion
stream hits the disk. 

The existence of azimuthal structure at the edge of the accretion
disks in LMXBs is empirically well-established, but theoretically
difficult to explain. Phenomenological modeling of the dipping sources
shows that this structure must have a scale height of $\sim$15\% of
the disk radius or more, and sustain itself over time intervals of
weeks to months.  However, such a scale height is more than can be
naturally produced by either irradiation or disk turbulence (King 1995
and references therein). On the other hand, material in ballistic
orbits cannot produce the heavily structured appearance of the
obscuring material. In the absence of a good physical model for the
vertical disk structure, the reason for its reduction or disappearance
can be only conjecture.  However, the disappearance may perhaps be
connected with a short-lived ($\sim$hrs-days) reduction in the
accretion flow from the secondary.  Such a variation could be smoothed
out by viscosity effects during the transfer of material through the
accretion disk, leaving the time-averaged X-ray flux unchanged.

\acknowledgments

This research has made use of data obtained through the High Energy
Astrophysics Science Archive Research Center Online Service, provided
by the NASA Goddard Space Flight Center.

\end{document}